\documentclass[bib]{statapress}

\usepackage{xcolor}
\hypersetup{
    colorlinks,
    linkcolor={red!50!black},
    citecolor={blue!50!black},
    urlcolor={blue!80!black}
}

\usepackage{my_sj}

\usepackage{epsfig}

\usepackage{stata}

\usepackage{amsmath, amsfonts}  

\usepackage{shadow}
 \textwidth15.75cm \evensidemargin5mm \oddsidemargin5mm
\newcommand{\jel}[1]{%
  \par%
  \noindent%
  {\small {\sffamily\bfseries JEL codes:} #1}%
}

\begin{document}

\inserttype[st0001]{article}
\author{Campos, Reggio, and Timini}{
Rodolfo G. Campos \\Banco de Espa\~na 
\and  
Iliana Reggio\\Universidad Aut\'onoma de Madrid
\and  
Jacopo Timini\\Banco de Espa\~na 
}  
\title[ge\_gravity2]{ge\_gravity2: a command for solving universal gravity models}

\maketitle

\begin{abstract} \noindent
We describe an algorithm for computing counterfactual trade flows, prices, output, and welfare in a large class of general equilibrium trade models. We introduce a command called \texttt{ge\_gravity2} that allows users to perform these computations in Stata. This command extends the existing \texttt{ge\_gravity} command by allowing users to compute the general equilibrium effects of changes in trade policy in positive supply elasticity models. It can be used to solve any model that falls into the class of universal gravity models as defined by Allen, Arkolakis, and Takahashi [Universal Gravity, \textit{Journal of Political Economy}, 128(2), 2020, 393--433].

\keywords{
general equilibrium, structural gravity, trade policy, \texttt{ge\_gravity2}}
\jel{F13, F17, D58}
\end{abstract}

\begin{center}
\end{center}




\section{Introduction}
In economics, the use of simulation models to analyze the impact of trade policies on trade flows is of great importance to policymakers and researchers alike. An early way of simulating structural gravity models in {\stata} was an ingenious procedure known as GEPPML due to \citet{AndersonLarchYotov:18}, which repurposed a Poisson pseudo-maximum-likelihood estimator to conduct general equilibrium simulations. \citet{Zylkin:19} later developed \texttt{ge\_gravity}, an independent command that can be used to simulate how trade flows and welfare respond to changes in bilateral trade costs, using the general equilibrium model and algorithm described by \citet{BayerYotovZylkin:19}. In this paper we introduce a new command called \texttt{ge\_gravity2} \citep{CamposReggioTimini:24}.

The \texttt{ge\_gravity2} command is designed to solve a more general class of models, since it applies to economic models that have a positive aggregate supply elasticity, whereas the \texttt{ge\_gravity} command is restricted to models in which the aggregate supply elasticity is zero. Another way in which the new command improves on the previous \texttt{ge\_gravity} command is that it can be used to simulate the effects of changes in technology parameters and trade balance parameters.

The \texttt{ge\_gravity2} command allows users to explore the effects of trade policy in trade models where the aggregate supply of goods is an increasing function of output prices and, more generally, to simulate a very wide range of economic geography models that combine aggregate demand and supply equations with standard market clearing conditions. This new command can be used to simulate the global effects of changes in trade frictions and supply shifters in any model that falls within the ``universal gravity'' framework described by \citet{AllenArkolakisTakahashi:20}.

A model belongs to the class of universal gravity models if an aggregate good is traded across locations and if it satisfies six specific economic conditions or properties. A model in this class must have (1) bilateral trade frictions of the ``iceberg'' type, (2) constant elasticity of substitution in aggregate demand and (3) constant elasticity of substitution in aggregate supply, (4) market clearing, (5) exogenous trade deficits, and (6) a choice of numeraire. These six conditions provide sufficient structure to fully describe all general equilibrium interactions of trade flows, incomes, and real output prices in terms of the elasticity of aggregate demand and supply.

The universal gravity framework encompasses a wide variety of models, including those of \citet{Armington:69}, \citet{Anderson:79}, \citet{AndersonvanWincoop:03}, \citet{Krugman:80}, \citet{EatonKortum:02}, \citet{Melitz:03}, \citet{diGiovanniLevchenko:13}, \citet{AllenArkolakis:14}, \citet{Redding:16}, and \citet{ReddingSturm:08}. 
On the other hand, many prominent models do not fall into the category of universal gravity models. These typically include models where there are multiple factors of production with varying intensities, or where trade deficits evolve endogenously, or where demand and supply elasticities are not constant. In addition, models that include additional sources of revenue, such as tariffs, generally do not fit into the universal gravity framework, except in special circumstances.

The paper is organized as follows. In Section~2, we present a prototypical trade model with positive supply elasticity, which serves as an example of the kind of models that can be simulated using the \texttt{ge\_gravity2} command. We then state the properties that characterize the universal gravity framework and show that the prototypical trade model belongs to this class of models. Next, we show how to derive the system of equations that can be used to compute comparative statics, and present an algorithm that solves for general equilibrium price changes for all universal gravity models. The system of equations (slightly) generalizes the one considered by \citet{AllenArkolakisTakahashi:20} in that it applies to unbalanced trade while allowing for changes in bilateral trade costs, exogenous trade costs, and parameters in the production function. The algorithm we present is fully global in the sense that it solves the nonlinear system of equations rather than a local approximation to that system. The algorithm uses a fixed-point procedure \`a la \citet{AlvarezLucas:07} that converges to the solution of the system of equations almost instantaneously in most applications.

In Section~3 we present the syntax of the \texttt{ge\_gravity2} command, the options that are available to the user, and the results stored by the command after its execution. Despite its broader scope, the usage of the \texttt{ge\_gravity2} command remains very similar to that of \texttt{ge\_gravity}, making it extremely easy for users of the previous command to seamlessly transition to the new command. The new command is backward compatible with \texttt{ge\_gravity} and will produce the same results as the previous command when the supply elasticity is set to zero. Compared to the current version of \texttt{ge\_gravity}, \texttt{ge\_gravity2} allows additional inputs and outputs and stores several matrices that can be inspected by the user after execution.

Section~4 demonstrates the use of the command through three examples. The first example demonstrates the basic use of the command by showing how to perform an ex ante analysis of the effects of a trade agreement. The second example demonstrates the use of the command with the \texttt{by} prefix by going through an example inspired by \citet{CamposReggioTimini:23} that calculates the evolution of welfare costs associated with trade policies in Spain during the Franco regime. The third example shows how the command can be used to simulate a scenario in which trade costs remain unchanged but a country's productivity increases, which in general equilibrium affects welfare and other economic variables for all countries in the world.

We conclude in Section~5 and show how to install the command in Section~7. In the Appendix, we derive various results that are used in the paper but are too long to include in the main text.

\section{Economic theory and methods}
\subsection{A prototypical trade model with a positive supply elasticity}
There are $N$ locations, denoted by the subscript $i$ or $j$. Sending goods from $i$ to $j$ incurs an iceberg trade cost, denoted by $\tau_{ij} \geq 1$. Goods are produced by combining immobile labor with intermediate inputs. As in the model of \citet{EatonKortum:02}, the production function is a Cobb-Douglas function with constant returns to scale, where $\zeta$ denotes the share of labor ($L_i$) in costs and $1 - \zeta$ the share of intermediate inputs ($M_i$), and $A_i > 0$ denotes labor productivity:
\begin{equation*}
    Q_i = (A_i L_i)^\zeta M_i^{1-\zeta}.
\end{equation*}

Intermediates are assumed to be the same bundle of goods as those entering final consumption, so that the price index for intermediates for each firm is the price index taken over all goods.\footnote{The assumption that the bundle of goods used in production and consumption is the same is often called roundabout production. It is a simple and tractable way to introduce a positive supply elasticity into a trade model. Although it may seem unrealistic, \citet{DhyneKikkawaMogstadTintelnot:21} conduct a careful empirical exercise using firm-level data from Belgium and conclude that what ultimately matters is how much the firm ultimately sells to foreign markets, not whether these sales are from direct or indirect exports. In an earlier version of their paper \citep{TintelnotDhyneKikkawaMogstad:18} they show that the quantitative results of a simple roundabout model like the one presented in this section are very close to the results of models with more complex production linkages.} We denote this price index by $P_i$.

Assuming perfect competition, the price of output at location $i$ is given by
\begin{equation*}
    p_i = \overline{\kappa} (w_i/A_i)^\zeta P_i^{1-\zeta},
\end{equation*}
where $\overline{\kappa} > 0$ is a constant and $w_i$ is the wage rate. The value of output at the origin is $Y_i \equiv p_i Q_i$. Since labor is the only factor of production and profits are zero, all output is distributed to workers. This leads to the following macroeconomic accounting identity:
\begin{equation*}
    Y_i = p_i Q_i = w_i L_i,
\end{equation*}
which means that $Y_i$ is simultaneously a measure of the value of production, labor income, and total income at a location.

Due to arbitrage in the goods markets, the price paid at a destination $j$ for a good that is shipped from the origin $i$ is
\begin{equation*}
    p_{ij} = \tau_{ij} p_i.
\end{equation*}

The amount of goods that reach the destination $j$ after subtracting the iceberg cost is denoted by $q_{ij}$. The expenditure on this good, valued at prices at the destination, is
\begin{equation*}
    X_{ij} = p_{ij} q_{ij}.
\end{equation*}

Equilibrium requires that output markets clear, that is, that prices and quantities adjust so that output $Q_i$ at each location equals the aggregate demand from all locations, including iceberg costs:
\begin{equation*}
    Q_i = \sum_{j=1}^N \tau_{ij} q_{ij}.
\end{equation*}

At each location there is a representative consumer who supplies labor inelastically and whose entire income comes from labor income. This consumer values varieties of goods according to a constant elasticity of substitution (CES) function that aggregates goods from all origins, as in the models of \citet{Armington:69}, \citet{Anderson:79}, and \citet{AndersonvanWincoop:03}. The elasticity of substitution is denoted by $\sigma>1$. The optimization problem of the consumer leads to the well-known result that expenditure on goods from different origins can be expressed as
\begin{equation*}
X_{ij} =  \frac{p_{ij}^{1-\sigma}}{\sum_{k=1}^N p_{kj}^{1-\sigma}} E_j = \frac{p_{ij}^{-\theta}}{\sum_{k=1}^N p_{kj}^{-\theta}} E_j,
\end{equation*}
where expenditure $E_j$ is defined by $E_j \equiv \sum_i X_{ij}$ and $\theta \equiv \sigma-1 > 0$. The parameter $\theta$ is known as the \textit{trade elasticity}.

Trade deficits are exogenous. Expenditure at any locations $i$ is expressed as a multiple of the value of output in the following way:
\begin{equation*}
    E_i = \Xi \xi_i p_i Q_i,
\end{equation*}
where
\begin{equation*}
    \Xi \equiv \frac{\sum_i p_i Q_i}{\sum_i \xi_i p_i Q_i}.
\end{equation*}
The parameter $\xi_i > 0$ is location specific and the parameter $\Xi$ is common to all locations. This way of specifying trade deficits ensures that the sum of exports from all locations always equals the sum of imports from all locations. In the special case where $\xi_i = 1$ for all $i$ (which also implies $\Xi = 1$), the trade deficits of all locations are exactly zero. On the other hand, if some $\xi_i \neq 1$, then trade is not balanced at all locations. The trade deficit at location $i$ can then be expressed as
\begin{equation*}
    D_{i} \equiv E_i - Y_i = (\Xi \xi_i - 1) p_i Q_i.
\end{equation*}
Summing trade deficits over all $i$, we have
\begin{equation*}
    \sum_{i=1}^N D_i = \Xi \left(\sum_{i=1}^N \xi_i p_i Q_i - \sum_{i=1}^N p_i Q_i\right) = \frac{\sum_i p_i Q_i}{\sum_i \xi_i p_i Q_i} \left(\sum_{i=1}^N \xi_i p_i Q_i - \sum_{i=1}^N p_i Q_i\right) = 0. 
\end{equation*}

In this model, output ($Q_i$), income ($Y_i$), and welfare ($W_i$) are proportional to real output prices ${p_i}/{P_i}$ raised to powers of the \textit{supply elasticity}, which is defined as $\psi \equiv (1-\zeta)/\zeta $.\footnote{We derive these results in the appendix. The online Appendix B of the article by \citet{AllenArkolakisTakahashi:20} derives a subset of these results. \citet{HeadMayer:22} derive similar results for welfare in what is essentially the same model, but solve the model in terms of real wages instead of real output prices.} The real output price ${p_i}/{P_i}$ is also a measure of terms of trade because it expresses the price obtained for exports of location $i$ in terms of prices paid for imports at location $i$ (including imports from itself). Using the notation ``$\propto$'' to express proportionality, we have that:
\begin{align*}
    Q_i \propto \frac{w_i}{p_i} L_i &\propto A_i L_i \left(\frac{p_i}{P_i}\right)^{\psi}, \\
    Y_i = p_i Q_i  &\propto A_i L_i\frac{p_i^{1+\psi}}{P_i^{\psi}},  \\
    W_i = \Xi \xi_i \frac{w_i}{P_i} &\propto \Xi \xi_i A_i \left(\frac{p_i}{P_i}\right)^{1+\psi}.
\end{align*}

The first of these results shows why $\psi$ is called a supply elasticity. Holding all other variables constant, the elasticity of output with respect to the price $p_i$ is given by
\begin{equation*}
    \frac{\partial \ln Q_i}{\partial \ln p_i} = \psi \geq 0.
\end{equation*}

The prototypical model described in this section falls into the more general class of models that \citet{AllenArkolakisTakahashi:20} define as the universal gravity framework. We now turn to the characterization of this more general framework and how it can be used to solve for the price ratio ${p_i}/{P_i}$, which---together with the supply elasticity $\psi$---is a sufficient statistic for real wages and ultimately welfare in the prototypical trade model.

\subsection{The universal gravity framework}

Models that are part of the universal gravity framework satisfy six properties. We list them briefly and refer the reader to the article by \citet{AllenArkolakisTakahashi:20} for a more detailed discussion.

\textbf{Property 1 (Arbitrage in goods markets)}. The bilateral price is equal to the product of the output price and a bilateral scalar (iceberg cost):
\begin{equation*}
    p_{ij} = \tau_{ij} p_i,
\end{equation*}
where $\tau_{ij} \in \mathbb{R}_{++} \cup \{\infty\}$ .

\textbf{Property 2 (CES aggregate demand)}. Expenditure at every location $j$ can be written as
\begin{equation*}
    E_j = \left( \sum_{i=1}^N p_{ij}^{-\theta}\right)^{-\frac{1}{\theta}} \mathcal{E}_j,
\end{equation*}
where $\mathcal{E}_j$ is real expenditure at location $j$ and the associated price index is $P_j \equiv \left(\sum_{i} p_{ij}^{-\theta}\right)^{-\frac{1}{\theta}}$. The parameter $\theta \in \mathbb{R}_{++}$ is called the \textit{demand elasticity} or \textit{trade elasticity}.

Because of Shephard's Lemma, Property~2 implies that the value of trade flows, defined as $X_{ij} \equiv p_{ij} q_{ij}$ (where $q_{ij}$ is the real value of goods arriving at location $j$ and $p_{ij}$ is their price at location $j$), can be written as a function of prices and expenditure as follows:
\begin{equation*}
X_{ij} = \frac{p_{ij}^{-\theta}}{\sum_{k} p_{kj}^{-\theta}} E_j.
\end{equation*}
Note that by summing this equation over $i$ we also obtain $\sum_i X_{ij} = E_j$.

\textbf{Property 3 (CES aggregate supply)}. Output at each location can be written as
\begin{equation*}
Q_i = \kappa \overline{c}_i \left(\frac{p_i}{P_i}\right)^\psi,
\end{equation*}
where $\kappa > 0$. The so-called supply shifters are strictly postive ($\overline{c}_i \in \mathbb{R}_{++}$) and the \textit{supply elasticity} $\psi$ is weakly positive ($\psi \in \mathbb{R}_{+}$).

\textbf{Property 4 (Output market clearing)}.
Output (supply) at each location equals demand from all locations, including iceberg trade costs:
\begin{equation*}
    Q_i = \sum_{j=1}^N \tau_{ij} q_{ij}.
\end{equation*}

\textbf{Property 5 (Exogenous deficits)}. For all $i$,
\begin{equation*}
    E_i = \Xi \xi_i p_i Q_i,
\end{equation*}
where
\begin{equation*}
    \Xi = \frac{\sum_i p_i Q_i}{\sum_i \xi_i p_i Q_i}.
\end{equation*}

\textbf{Property 6 (Price normalization)}. 
    \begin{equation*}
        \sum_{i=1}^N Y_i \equiv \sum_{i=1}^N p_i Q_i = \overline{Y} > 0.
    \end{equation*}
This last property is stated by \citet{AllenArkolakisTakahashi:20} for the case with $\overline{Y} = 1$, but it will be more convenient for the implementation of the algorithm to use a different constant because it avoids having to re-scale the data before and after the algorithm.

The consumption side of the prototypical trade model described in the previous section satisfies Property~2, and the production side satisfies Property~3 with $A_i L_i = \overline{c}_i$. Properties 1,~4, and~5 are stated directly in the description of this model. Finally, Property~6 is a normalization that does not conflict with the prototypical model because an equilibrium in that model determines the value of relative prices but not the price level, which is a free parameter. This additional degree of freedom is removed by the normalization.

\citet{AllenArkolakisTakahashi:20} show that various other models also satisfy the six properties of the universal gravity framework, among them the models by \citet{Armington:69}, \citet{Anderson:79}, \citet{AndersonvanWincoop:03}, \citet{Krugman:80}, \citet{EatonKortum:02}, \citet{Melitz:03}, \citet{diGiovanniLevchenko:13}, \citet{AllenArkolakis:14}, \citet{Redding:16}, and \citet{ReddingSturm:08}.

\subsection{Comparative statics}

\citet{AllenArkolakisTakahashi:20} show that the response of prices to a change in the trade cost matrix can be obtained by solving a system of equations that is the same for all models in the universal gravity framework that have the same trade elasticity $\theta$ and supply elasticity $\psi$. In this section we use a slightly more general version of their system of equations that allows for unbalanced trade, changes in trade deficits, and changes in supply shifters.

We use the ``hat algebra'' notation introduced by \citet{DekleEatonKortum:08}. For each variable, its hat version is defined as the ratio of the value of the variable in a counterfactual equilibrium relative to the value in a baseline equilibrium; if $x$ is the value in the baseline equilibrium and $x'$ is the value in the counterfactual equilibrium, then $\hat{x} \equiv x'/x$.

The comparative statics exercise we consider allows for a change in the matrix of bilateral trade costs with elements $\hat{\tau}_{ij} = \tau'_{ij}/\tau_{ij}$ and also for changes in the vectors of parameters $\hat{\xi}_i = \xi'_i/\xi_i$, which affect trade deficits, and in the supply shifters $\hat{c}_i = \overline{c}'_i/\overline{c}_i$. The bilateral trade flow matrix $\mathbf{X} = [X_{ij}]_{N \times N}$ is assumed to be known. Income and expenditure by location, and the global value of trade are obtained from the bilateral trade flow matrix as $Y_i \equiv \sum_j X_{ij}$, $E_i \equiv \sum_j X_{ji}$, and $\overline{Y} \equiv \sum_i \sum_j X_{ij}$.

We show in the appendix that if prices and quantities in the baseline and the counterfactual are part of an equilibrium that satisfies the six properties of the universal gravity framework, then the changes in price vectors $(\hat{p}_i)_{N \times 1}$ and $(\hat{P}_i)_{N \times 1}$ must solve the following system of $2N+1$ nonlinear equations:
\begin{align*}
    \hat{p}_i^{1+\theta+\psi} \hat{P}_i^{-\psi} \hat{c}_i &= \hat{\Xi}\sum_j \left[\frac{X_{ij}}{Y_i}\right] (\hat{\tau}_{ij})^{-\theta} (\hat{P}_j)^\theta \hat{p}_j  \hat{\xi}_j \hat{c}_j \left( \frac{\hat{p}_j}{\hat{P}_j} \right)^\psi, \quad i=1, \dots, N \\
    \hat{P}_i^{-\theta} &= \sum_{j} \left[ \frac{X_{ji}}{E_i} \right] \hat{\tau}_{ji}^{-\theta} \hat{p}_{j}^{-\theta}, \quad i=1, \dots, N, \\
    \hat{\Xi} &= \frac{1}{\sum_{i=1}^N \hat{\xi}_i \hat{c}_i\frac{\hat{p_i}^{1+\psi}}{\hat{P}_i^\psi} (E_i/\overline{Y})}.
\end{align*}

Once this system is solved, other quantities can be derived from the resulting price vectors. For example, in the case of the prototypical trade model, because of the proportionality relationships shown earlier, the comparative statics for output and income can be obtained immediately as
\begin{align*}
    \hat{Q}_i &= \frac{\hat{w}_i}{\hat{p}_i} = \hat{c}_i \left(\frac{\hat{p}_i}{\hat{P}_i}\right)^{\psi}, \\
    \hat{Y}_i &= \hat{p}_i \hat{Q}_i  = \hat{c}_i \frac{\hat{p}_i^{1+\psi}}{\hat{P}_i^{\psi}},
\end{align*}
where $\hat{c}_i = \hat{A}_i \hat{L}_i$, and the comparative of statics for welfare are given by
\begin{equation*}
    \hat{W}_i = \hat{\Xi} \hat{\xi}_i \frac{\hat{w}_i}{\hat{P}_i} = \hat{\Xi} \hat{\xi}_i \frac{\hat{c}_i}{\hat{L}_i} \left(\frac{\hat{p}_i}{\hat{P}_i}\right)^{1+\psi}.
\end{equation*}
Comparative statics for expenditure are determined by
\begin{equation*}
    \hat{E}_i = \hat{\Xi} \hat{\xi}_i \hat{Y}_i,
\end{equation*}
and the change in bilateral trade flows trade flows can be calculated as
\begin{equation*}
    \hat{X}_{ij} = \hat{\tau}_{i, j}^{-\theta} \hat{p}_i^{-\theta} \hat{P}_j^\theta  \hat{E}_j.
\end{equation*}

The comparative statics derived for welfare and output are specific to the prototypical model presented in this article. Other models may yield different results, especially for welfare. However, the comparative statics for some of the variables are quite general in that they are common to all models in the universal gravity framework. This holds for prices and, conditional on the price normalization, also for the value bilateral trade flows, expenditure, and income.

\subsection{Algorithm to solve for prices (universal option)}

The algorithm used to solve for prices is a fixed-point algorithm in the style of \citet{AlvarezLucas:07}. Its structure and implementation is very similar to the algorithm used by \citet{Zylkin:19} in the \texttt{ge\_gravity} command. There are two main differences. The first difference is that the algorithm solves for the vector of output prices instead of wages. The reason for this is that the universal gravity framework is agnostic about which factors of production are involved in the production process. In fact, there may be models within the universal gravity framework for which there is no properly defined wage rate. The second difference is that the algorithm has more inputs than the algorithm in \texttt{ge\_gravity}, since it takes into account exogenous changes in supply shifters and arbitrary exogenous adjustments to trade deficits in addition to the value of the supply elasticity.

\textbf{Notation}. In the description of the algorithm, we use the commands \texttt{rowsum}, \texttt{colsum}, and \texttt{sum}, which are defined as follows (this is how they work in Mata). The command \texttt{rowsum} sums the elements of a matrix along each row and returns a column vector. The command \texttt{colsum} sums the elements of a matrix along each column and returns a row vector. The command \texttt{sum} sums all elements of a vector or matrix and returns a scalar. The symbol $\otimes$ represents the Kronecker product. All other operations, such as multiplication, division, and exponentiation, are performed element by element. All vectors are $N \times 1$ column vectors and all matrices are $N \times N$ square matrices, where $N$ is the number of locations. We use the notation $\mathbf{1}$ for a column vector of size $N \times 1$ filled with ones. The operation of transposing a vector or matrix is denoted by the superscript $T$.

\textbf{Inputs}. The required inputs to the algorithm consist of a matrix of bilateral trade flows, which we will call $\mathbf{X}$, a matrix $\mathbf{B}$ that encodes the change in bilateral trade costs, and two scalars, one for the trade elasticity and one for the supply elasticity.

The matrix $\mathbf{X}$ is a square matrix containing bilateral trade flows from locations in the rows to locations in the columns. The order of the locations that appear in the rows and columns must match, and there must be no missing values (although there can be zeros). This matrix is created from a user-supplied variable containing bilateral trade flows. 

The second key input is a matrix that encodes the partial equilibrium effects of changes in trade costs. We will call this matrix $\mathbf{B}$. It has the same dimensions as $\mathbf{X}$ and the element in row $i$ and column $j$ codifies how trade costs would affect trade flows from location $i$ to location $j$ in partial equilibrium. In terms of the notation used in the model, the element in row $i$ and column $j$ of $\mathbf{B}$ corresponds to $\tau_{ij}^{-\theta}$.\footnote{This matrix is internal to the command and is constructed from a variable provided by the user. The examples show how to create this variable in {\stata}.} 

The remaining two required inputs are the trade elasticity $\theta$ and the supply elasticity $\psi$. In addition to the required inputs, there are two additional input vectors that can be optionally provided to the algorithm: the vector $\mathbf{\hat{\xi}}$, which governs the discrepancy between expenditures and revenues, and ultimately trade deficits, and the vector $\mathbf{\hat{c}}$, which governs the relative changes in the supply shifters. If these optional inputs are not provided, then they are set to be column vectors of size $N \times 1$ filled with ones.

\textbf{Algorithm (universal option)}. The algorithm solves the system of equations for prices by iterating until the vector of output prices converges. The steps involved in the iteration are as follows:

\textbf{Step 0}. Choose a tolerance level for convergence $\varepsilon>0$. Compute the vectors $\mathbf{E} = \texttt{colsum}(\mathbf{X})^T$, $\mathbf{Y} = \texttt{rowsum}(\mathbf{X})$, and the scalar $\overline{Y} = \texttt{sum}(\mathbf{X})$. Initialize the vectors $\mathbf{\hat{p}} = \mathbf{1}$ and $\mathbf{\hat{P}} = \mathbf{1}$. 

\textbf{Step 1}. Compute the scalar $\hat{\Xi}$:
\begin{equation*}
    \hat{\Xi} = \overline{Y} / \texttt{sum} \left(\mathbf{\hat{\xi}}\,\mathbf{\hat{c}} \,\hat{\mathbf{p}}\left(\frac{\hat{\mathbf{p}}}{\hat{\mathbf{P}}}\right)^\psi\mathbf{E}\right).
\end{equation*}

\textbf{Step 2}. Update the vector $\mathbf{\hat{p}}$:
\begin{equation*}
    \mathbf{\hat{p}}_{\text{next}} =  \left(\hat{\Xi}\left(\frac{\mathbf{\hat{P}}^{\psi}}{\mathbf{\hat{c}}}\right) \, \texttt{rowsum}\left(\left[\frac{\mathbf{X}}{\mathbf{Y} \otimes \mathbf{1}^T}\right] \mathbf{B} \left[ \left(\mathbf{\hat{\xi}}\,\mathbf{\hat{c}}\,\mathbf{\hat{P}}^{\theta-\psi} \mathbf{\hat{p}}^{1+\psi}\right)^T \otimes \mathbf{1} \right]\right)\right)^\frac{1}{1+\theta+\psi}
\end{equation*}

\textbf{Step 3}. Update the vector $\mathbf{\hat{P}}$:
\begin{equation*}
 \mathbf{\hat{P}}_{\text{next}} =  \left(\texttt{colsum}\left(\left[\frac{\mathbf{X}}{\mathbf{E}^T \otimes \mathbf{1}}\right] \mathbf{B} \left[ \mathbf{\hat{p}}_{\text{next}}^{-\theta} \otimes  \mathbf{1}^T \right] \right)^T\right)^{-\frac{1}{\theta}}
\end{equation*}

\textbf{Step 4}. Check convergence:
\begin{equation*}
    \lVert \mathbf{\hat{p}}_{\text{next}} - \mathbf{\hat{p}} \rVert_{\sup} < \varepsilon
\end{equation*}
If convergence was achieved, then stop. If not, then set $\mathbf{\hat{p}} = \mathbf{\hat{p}}_{next}$ and $\mathbf{\hat{P}} = \mathbf{\hat{P}}_{next}$ and return to step 1.

\textbf{Outputs}. Once the algorithm has converged, the vector of normalized output prices $\hat{\mathbf{p}}$, the vector of normalized price indices $\hat{\mathbf{P}}$, and the scalar $\hat{\Xi}$ are known.

\subsection{Algorithm to solve for prices with constant trade deficits (default option)}
The algorithm just described is the one used when the user runs the \texttt{ge\_gravity2} command with the \texttt{universal} option.
For many applications, it may be more convenient to assume that trade deficits remain constant instead of varying with income. The case with constant trade deficits is the default option of the existing \texttt{ge\_gravity} command, and we make it the default of the newer \texttt{ge\_gravity2} command as well. This case can be implemented as a special case in the universal gravity framework by making the change in the $\xi_i$ parameters endogenous.

To show this, we start from the equation
\begin{equation*}
    \Xi \xi_i p_i Q_i = E_i = D_i + p_i Q_i.
\end{equation*}
The equality on the left is given as a definition in Property~5 of the universal gravity framework, and the equality on the right is the usual definition of trade deficits.

Solving this equation for $\xi_i$ leads to
\begin{equation*}
    \xi_i = \frac{1}{\Xi} \left(1 + \frac{D_i}{p_i Q_i}\right) = \frac{1}{\Xi} \left(1 + \delta_i\right),
\end{equation*}
where we have defined $\delta_i$ as the ratio of the trade deficit to income:
\begin{equation*}
    \delta_i \equiv \frac{D_i}{p_i Q_i}.
\end{equation*}

The change in the endogenous variable $\hat{\xi}_i$ is then determined by 
\begin{equation*}
    \hat{\xi}_i = \frac{1}{\hat{\Xi}} \frac{1 + \delta'_i}{1+\delta_i}  = \frac{1}{\hat{\Xi}} \frac{1 + \hat{\delta}_i\delta_i}{1+\delta_i} = \frac{1}{\hat{\Xi}} \left( 1 + \frac{\delta_i}{1+\delta_i} (\hat{\delta_i}-1)\right).
\end{equation*}
The last step is to derive an expression for $\hat{\delta_i}$ in terms of price changes and supply shifters. Using the condition that deficits remain constant, i.e., $D'_i = D_i$, this can be done as follows:
\begin{equation*}
    \hat{\delta_i} = \frac{D_i/(p'_i Q'_i)}{D_i/(p_i Q_i)} = \frac{1}{\hat{p}_i \hat{Q}_i} = \frac{1}{\hat{p}_i \hat{Q}_i} = \frac{1}{\hat{Y}_i} = \frac{\hat{P}_i^\psi}{\hat{c}_i \hat{p}_i^{1+\psi}}
\end{equation*}
Finally, putting the previous results together, the endogenous change of $\hat{\xi}_i$ required to maintain trade deficits constant is
\begin{equation*}
    \hat{\xi}_i = \frac{1}{\hat{\Xi}} \left( 1 + \frac{\delta_i}{1+\delta_i} \left(\frac{\hat{P}_i^\psi}{\hat{c}_i \hat{p}_i^{1+\psi}}-1\right)\right).
\end{equation*}

The algorithm that \texttt{ge\_gravity2} uses by default is a modified version of the algorithm for the universal case. The modifications consist of initializing $\hat{\Xi}$ to one and adding an extra step just before step 1 that calculates the endogenous change in $\hat{\xi}_i$ parameters that keeps trade deficits constant.

\textbf{Algorithm (default option)}.

\textbf{Step 0}. Choose a tolerance level for convergence $\varepsilon>0$. Compute the vectors $\mathbf{E} = \texttt{colsum}(\mathbf{X})^T$, $\mathbf{Y} = \texttt{rowsum}(\mathbf{X})$, and the scalar $\overline{Y} = \texttt{sum}(\mathbf{X})$. Compute the vector $\mathbf{\delta} = (\mathbf{E}-\mathbf{Y})/\mathbf{Y}$.  Initialize the vectors $\mathbf{\hat{p}} = \mathbf{1}$ and $\mathbf{\hat{P}} = \mathbf{1}$. Initialize the scalar $\hat{\Xi} = 1$.

\textbf{Step 0.5}. Compute the vector $\mathbf{\hat{\xi}}$:
\begin{equation*}
    \mathbf{\hat{\xi}} = \frac{1}{\hat{\Xi}} \left( \mathbf{1} + \frac{\mathbf{\delta}}{\mathbf{1}+\mathbf{\delta}} \left(\frac{\mathbf{\hat{P}}^\psi}{\mathbf{\hat{c}}\, \mathbf{\hat{p}}^{1+\psi}}-\mathbf{1}\right)\right).
\end{equation*}

\textbf{Step 1}. Compute the scalar $\hat{\Xi}$:
\begin{equation*}
    \hat{\Xi} = \overline{Y} / \texttt{sum} \left(\mathbf{\hat{\xi}}\,\mathbf{\hat{c}} \,\hat{\mathbf{p}}\left(\frac{\hat{\mathbf{p}}}{\hat{\mathbf{P}}}\right)^\psi\mathbf{E}\right).
\end{equation*}

\textbf{Step 2}. Update the vector $\mathbf{\hat{p}}$:
\begin{equation*}
    \mathbf{\hat{p}}_{\text{next}} =  \left(\hat{\Xi}\left(\frac{\mathbf{\hat{P}}^{\psi}}{\mathbf{\hat{c}}}\right) \, \texttt{rowsum}\left(\left[\frac{\mathbf{X}}{\mathbf{Y} \otimes \mathbf{1}^T}\right] \mathbf{B} \left[ \left(\mathbf{\hat{\xi}}\,\mathbf{\hat{c}}\,\mathbf{\hat{P}}^{\theta-\psi} \mathbf{\hat{p}}^{1+\psi}\right)^T \otimes \mathbf{1} \right]\right)\right)^\frac{1}{1+\theta+\psi}
\end{equation*}

\textbf{Step 3}. Update the vector $\mathbf{\hat{P}}$:
\begin{equation*}
 \mathbf{\hat{P}}_{\text{next}} =  \left(\texttt{colsum}\left(\left[\frac{\mathbf{X}}{\mathbf{E}^T \otimes \mathbf{1}}\right] \mathbf{B} \left[ \mathbf{\hat{p}}_{\text{next}}^{-\theta} \otimes  \mathbf{1}^T \right] \right)^T\right)^{-\frac{1}{\theta}}
\end{equation*}

\textbf{Step 4}. Check convergence:
\begin{equation*}
    \lVert \mathbf{\hat{p}}_{\text{next}} - \mathbf{\hat{p}} \rVert_{\sup} < \varepsilon
\end{equation*}
If convergence was achieved, then stop. If not, then set $\mathbf{\hat{p}} = \mathbf{\hat{p}}_{next}$ and $\mathbf{\hat{P}} = \mathbf{\hat{P}}_{next}$ and return to step~0.5.

\textbf{Outputs}. Once the algorithm has converged, the vector of normalized output prices $\hat{\mathbf{p}}$, the vector of normalized price indices $\hat{\mathbf{P}}$, and the scalar $\hat{\Xi}$ are known.

Running the \texttt{ge\_gravity2} command with a supply elasticity ($\psi$) of zero and no other options will replicate the results of \texttt{ge\_gravity} without any other options.

\subsection{The multiplicative option}
The \texttt{ge\_gravity} command has a \texttt{multiplicative} option in which trade deficits are assumed to evolve at all locations in a way that satisfies the equation
\begin{equation*}
    \hat{E}_i = \frac{Y'_i + D'_i}{Y_i + D_i} = \hat{Y}_i.
\end{equation*}

In \texttt{ge\_gravity2} we also implement the \texttt{multiplicative} option to make the command backward compatible with \texttt{ge\_gravity}. This option does not require a different solution algorithm. The algorithm for the \texttt{universal} option is used, setting $\hat{\xi}_i$ = 1 for all $i$. After the algorithm has converged, the value of the constant $\hat{\Xi}$ is changed to one. This rescales expenditures to enforce the equality $\hat{E}_i = \hat{Y}_i$. Running the \texttt{ge\_gravity2} command with the \texttt{multiplicative} option and a supply elasticity ($\psi$) of zero will replicate the results obtained by \texttt{ge\_gravity} with the \texttt{multiplicative} option.

\subsection{Remaining calculations}
Once price changes have been obtained, other variables follow in order. We maintain the convention that all operations, except the Kronecker product, are performed element by element. The change in income is obtained directly from the price changes as
\begin{equation*}
    \hat{\mathbf{Y}} = \mathbf{\hat{c}} \,\hat{\mathbf{p}}\left(\frac{\hat{\mathbf{p}}}{\hat{\mathbf{P}}}\right)^\psi.
\end{equation*}
The change in expenditure then follows as
\begin{equation*}
    \hat{\mathbf{E}} = \hat{\Xi}\, \hat{\mathbf{\xi}}\,\hat{\mathbf{Y}}.
\end{equation*}
Changes in bilateral trade flows are obtained as
\begin{equation*}
    \hat{\mathbf{X}} = \mathbf{B} \left[\frac{\hat{\mathbf{p}} \otimes \mathbf{1}^T}{\hat{\mathbf{P}}^T \otimes \mathbf{1}}\right]^{-\theta}\left[\hat{\mathbf{E}}^T \otimes \mathbf{1} \right].
\end{equation*}

For the prototypical trade model, the change in real prices also translates directly into changes in output, which is obtained as
\begin{equation*}
    \hat{\mathbf{Q}} = \mathbf{\hat{c}} \left(\frac{\hat{\mathbf{p}}}{\hat{\mathbf{P}}}\right)^{\psi}.
\end{equation*}
The computation for the change in welfare depends on whether the change in supply shifters is generated by changes in productivity or the labor force. For the prototypical trade model, the vector $\mathbf{\hat{c}}$ is the product of $\mathbf{\hat{A}}$ and $\mathbf{\hat{L}}$, and the change in welfare can be expressed as 
\begin{equation*}
    \hat{\mathbf{W}} = \hat{\Xi}\, \hat{\mathbf{\xi}}\,\frac{\mathbf{\hat{c}}}{\mathbf{\hat{L}}} \left(\frac{\hat{\mathbf{p}}}{\hat{\mathbf{P}}}\right)^{1+\psi} = \hat{\Xi}\, \hat{\mathbf{\xi}}\,\mathbf{\hat{A}} \left(\frac{\hat{\mathbf{p}}}{\hat{\mathbf{P}}}\right)^{1+\psi}.
\end{equation*}
The syntax of the \texttt{ge\_gravity2} command includes options that allow the user to specify $\mathbf{\hat{A}}$ and $\mathbf{\hat{L}}$, or to specify $\mathbf{\hat{c}}$. Welfare is calculated only if the user specifies $\mathbf{\hat{A}}$ or $\mathbf{\hat{L}}$, or both at the same time, but not if the user specifies $\mathbf{\hat{c}}$.

The changes of real and nominal wages in the prototypical trade model also depend on the distinction between $\mathbf{\hat{A}}$ and $\mathbf{\hat{L}}$. They are calculated as
\begin{equation*}
    \frac{\hat{\mathbf{w}}}{\hat{\mathbf{P}}} = \mathbf{\hat{A}} \left(\frac{\hat{\mathbf{p}}}{\hat{\mathbf{P}}}\right)^{1+\psi}
\end{equation*}
and
\begin{equation*}
    \hat{\mathbf{w}} = \hat{\mathbf{P}} \mathbf{\hat{A}} \left(\frac{\hat{\mathbf{p}}}{\hat{\mathbf{P}}}\right)^{1+\psi} =  \mathbf{\hat{A}} \frac{\hat{\mathbf{p}}^{1+\psi}}{\hat{\mathbf{P}}^{\psi}}.
\end{equation*}

\subsection{Reporting comparative statics for the prototypical trade model}
The normalization in Property 6 determines the scale of nominal quantities in all universal gravity models. Since the prototypical trade model determines only real variables, we said earlier that Property~6 does not conflict with the model. However, there is a subtle point that needs to be clarified. The use of normalization in Property~6 is harmless for equilibrium objects in the baseline economy and in the counterfactual economy when considered in isolation. However, when doing comparative statics, using the \textit{same} normalization in \textit{both} models introduces an additional assumption. In this case, the assumption is that global nominal income is the same in the baseline and the counterfactual. This assumption may not be consistent with the intended use of the model. Researchers should therefore be cautious when reporting comparative static calculations for nominal quantities. For this reason, the default results table generated by the command shows growth rates only for real variables.

The default calculation reports the vector relative changes of real (non-domestic) exports for all locations, which is calculated as
\begin{equation*}
    \mathbf{g}_{\,exp} = 100 \times \left(\left(\frac{\texttt{rowsum}\left(\mathbf{\hat{X}}\mathbf{X} \,(\mathbf{1}_{N \times N} -  \mathbf{I}_{N \times N})\right)}{ \texttt{rowsum}\left(\mathbf{X} \, (\mathbf{1}_{N \times N} -  \mathbf{I}_{N \times N})\right)}\right) \Big/ \mathbf{\hat{p}} - \mathbf{1}_{N \times 1} \right)
\end{equation*}
Here, $\mathbf{1}_{N \times N}$ denotes a square matrix of size $N$ filled with ones and $\mathbf{1}_{N \times 1}$ denotes a $N \times 1$ unitary vector. We use the notation $\mathbf{I}_{N \times N}$ to refer to the identity matrix of size $N$. We continue using the convention that all operations are performed element by element. Notice that multiplying a matrix by the term $(\mathbf{1}_{N \times N} -  \mathbf{I}_{N \times N})$ is equivalent to setting its diagonal elements to zero. We do this to exclude domestic trade from the calculation. The relative change in exports is real (as opposed to nominal) because the nominal change of exports is deflated by the change in export prices $\mathbf{\hat{p}}$ in the equation above.

Similarly, the vector of real relative changes of imports is calculated as
\begin{equation*}
    \mathbf{g}_{\,imp} = 100 \times \left(\left(\frac{\texttt{colsum}\left(\mathbf{\hat{X}}\mathbf{X} \,(\mathbf{1}_{N \times N} -  \mathbf{I}_{N \times N})\right)}{ \texttt{colsum}\left(\mathbf{X} \, (\mathbf{1}_{N \times N} -  \mathbf{I}_{N \times N})\right)}\right)^T \Big/ \mathbf{\hat{P}} - \mathbf{1}_{N \times 1} \right)
\end{equation*}
This expression is very similar to the expression for exports. The main differences are that matrices are summed over columns (i.e., across rows for each column) to obtain imports instead of exports, and that the nominal change is deflated by the change in the consumer price index $\mathbf{\hat{P}}$, instead of the change in export prices $\mathbf{\hat{p}}$.

The relative change in total trade is obtained as the weighted average of the relative changes of exports and imports, where the weights are given by exports and imports in the baseline economy. The change in domestic trade is calculated from the diagonal elements of the bilateral trade matrix $\mathbf{\hat{X}}$ as follows:
\begin{equation*}
    \mathbf{g}_{\,dom} = 100 \times \left(\texttt{diagonal}\left(\mathbf{\hat{X}}\right) / \mathbf{\hat{P}} - \mathbf{1}_{N \times 1} \right).
\end{equation*}
This implies that domestic trade is also deflated by the consumer price index $\mathbf{\hat{P}}$.

Finally, the relative change of output and welfare are computed as
\begin{equation*}
    \mathbf{g}_{\,Q} = 100 \times \left(\mathbf{\hat{Q}} - \mathbf{1}_{N \times 1} \right)
\end{equation*}
and
\begin{equation*}
    \mathbf{g}_{\,W} = 100 \times \left(\mathbf{\hat{W}} - \mathbf{1}_{N \times 1} \right).
\end{equation*}
All these relative changes are expressed in percentage points. Thay are to be interpreted as the impact of moving from the baseline situation to the counterfactual situation, not as growth rates of variables over time.

\section{The \texttt{ge\_gravity2} command}
\subsection{Syntax}


\begin{stsyntax}
\texttt{ge\_gravity2}
    {\tt exp\_id}  {\tt imp\_id} {\tt flows} {\tt partial}
    \optif
    \optin, theta()
    \optional{{\tt psi() gen\_options other\_options}}
\end{stsyntax}

\textbf{Required variables}
\begin{itemize}
    \item \texttt{exp\_id} specifies the variable that identifies the location of origin, for example
    ISO codes or names of countries. This variable can be a string variable or numeric. This variable cannot contain missing values.
    \item \texttt{imp\_id} specifies the variable that identifies the location of destination, for example ISO codes or names of countries. This variable a string variable or numeric. This variable cannot contain missing values.
    \item \texttt{flows} contains bilateral trade flows. This variable contains the flow from the location identified as \texttt{exp\_id} to the location identified as \texttt{imp\_id}. This variable cannot contain missing values, but can contain zeros.
    \item \texttt{partial} contains the ``partial'' estimate of the effect, typically obtained as a coefficient from a prior gravity estimation. This variable cannot contain missing values.
\end{itemize}

\textbf{Elasticities}
\begin{itemize}
    \item \texttt{theta(\#)} sets the trade elasticity, which must be strictly positive.
    \item \texttt{psi(\#)} sets the aggregate supply elasticity, which must be nonnegative. If the option \texttt{psi()} is not specified, then the default is \texttt{psi(0)}. With a supply elasticity of zero, the command solves essentially the same model as \texttt{ge\_gravity}.
\end{itemize}

\subsection{Options}

\textbf{New variables (universal gravity)}
\begin{itemize}
    \item \texttt{gen\_X}(\textit{varname}) generates counterfactual trade flows ($\mathbf{X}'$ in the model) and places the result in a new variable called \textit{varname}, or overwrites this variable if it already exists.
    \item \texttt{gen\_rp}(\textit{varname}) generates the change in real prices ($\hat{\mathbf{p}} / \hat{\mathbf{P}}$ in the model) and places the result in a new variable called \textit{varname}, or overwrites this variable if it already exists.
    \item \texttt{gen\_y}(\textit{varname}) generates the change in income ($\hat{\mathbf{Y}}$ in the model) and places the result in a new variable called \textit{varname}, or overwrites this variable if it already exists.
    \item \texttt{gen\_x}(\textit{varname}) generates the change in trade flows ($\hat{\mathbf{X}}$ in the model) and places the result in a new variable called \textit{varname}, or overwrites this variable if it already exists.
    \item \texttt{gen\_p}(\textit{varname}) generates the change in output prices ($\hat{\mathbf{p}}$ in the model) and places the result in a new variable called \textit{varname}, or overwrites this variable if it already exists.
    \item \texttt{gen\_P}(\textit{varname}) generates the change in price indices ($\hat{\mathbf{P}}$ in the model) and places the result in a new variable called \textit{varname}, or overwrites this variable if it already exists.
\end{itemize}

\textbf{New variables (prototypical model)}
\begin{itemize}
    \item \texttt{gen\_w}(\textit{varname}) generates the change in welfare ($\hat{\mathbf{W}}$ in the model) and places the result in a new variable called \textit{varname}, or overwrites this variable if it already exists.
    \item \texttt{gen\_q}(\textit{varname}) generates the change in output ($\hat{\mathbf{Q}}$ in the model) and places the result in a new variable called \textit{varname}, or overwrites this variable if it already exists.
    \item \texttt{gen\_rw}(\textit{varname}) generates the change in real wages ($\hat{\mathbf{w}}/\hat{\mathbf{P}}$ in the model) and places the result in a new variable called \textit{varname}, or overwrites this variable if it already exists.
    \item \texttt{gen\_nw}(\textit{varname}) generates the change in nominal wages ($\hat{\mathbf{w}}$ in the model) and places the result in a new variable called \textit{varname}, or overwrites this variable if it already exists.
\end{itemize}

\textbf{Other options}
\begin{itemize}
    \item \texttt{\underline{r}esults} prints a table with percent changes for exports, imports, total trade, domestic trade, output, and welfare.
    \item  \texttt{\underline{uni}versal} solves the model with universal trade deficits and allows the user to set the option \texttt{xi\_hat}.
    \item  \texttt{\underline{mult}iplicative} solves the model for trade deficits that imply $\hat{\mathbf{E}} = \hat{\mathbf{Y}}$ (for backward compatibility with the multiplicative option of the \texttt{ge\_gravity} command).
    \item  \texttt{c\_hat(matrix)} changes supply shifters ($\mathbf{\hat{c}}$ in the model). Welfare will not be calculated if this option is used. This option may not be combined with the options \texttt{a\_hat} or \texttt{l\_hat}. The default behavior is that all elements of \texttt{c\_hat} are set to one.
    \item  \texttt{a\_hat(matrix)} changes productivity ($\mathbf{\hat{A}}$ in the prototypical trade model). This option may not be combined with the option \texttt{c\_hat}. The default behavior is that all elements of \texttt{a\_hat} are set to one.
    \item  \texttt{l\_hat(matrix)} changes the labor force ($\mathbf{\hat{L}}$ in the prototypical trade model). This option may not be combined with the option \texttt{c\_hat}. The default behavior is that all elements of \texttt{l\_hat} are set to one.
    \item \texttt{xi\_hat(matrix)} changes trade deficits  ($\mathbf{\hat{\xi}}$ in the model). This option must be used in combination with the \texttt{universal} option. If the \texttt{universal} option is selected and the \texttt{xi\_ha}t option is not used, then the command defaults to setting all elements of \texttt{xi\_hat} to one.
    \item \texttt{tol(\#)} sets the tolerance level to verify convergence of the vector of output price changes. The tolerance level must be a strictly positive real number. The default is \texttt{tol(1e-12)}.
    \item \texttt{max\_iter(\#)} sets the maximum number of iterations to solve for the vector of output price changes. The maximum number of iterations must be a positive integer. The default is \texttt{max\_iter(1000000)}.
\end{itemize}

\subsection{Stored results}

\textbf{Scalars}
\begin{itemize}
    \item \texttt{e(theta)}: trade elasticity
    \item \texttt{e(psi)}: supply elasticity
    \item \texttt{e(N)}: number of locations
    \item \texttt{e(crit)}: convergence criterion achieved
    \item \texttt{e(n\_iter)}: number of iterations performed
    \item \texttt{e(Xi\_hat)}: scalar to ensure that trade deficits sum to zero
\end{itemize}

\textbf{Macros}
\begin{itemize}
    \item \texttt{e(names)}: identifiers of locations
\end{itemize}

\textbf{Matrices}
\begin{itemize}
    \item \texttt{e(results)}: $N \times 6$ matrix of relative changes for key variables
    \item \texttt{e(X)}: $N \times N$ matrix containing baseline trade flows $\mathbf{X}$
    \item \texttt{e(X\_hat)}: $N \times N$ matrix containing $\hat{\mathbf{X}} = \mathbf{X}' / \mathbf{X}$
    \item \texttt{e(X\_prime)}: $N \times N$ matrix containing counterfactual trade flows $\mathbf{X}'$
    \item \texttt{e(E)}: $N \times 1$ matrix containing baseline expenditure $\mathbf{E}$
    \item \texttt{e(Y)}: $N \times 1$ matrix containing baseline income $\mathbf{Y}$
    \item \texttt{e(rp)}: $N \times 1$ matrix containing $\hat{\mathbf{p}} / \hat{\mathbf{P}}$
    \item \texttt{e(p\_hat)}: $N \times 1$ matrix of the change in output prices ($\hat{\mathbf{p}}$)
    \item \texttt{e(P\_hat)}: $N \times 1$ matrix of the change in price indices ($\hat{\mathbf{P}}$)
    \item \texttt{e(E\_hat)}: $N \times 1$ matrix containing $\hat{\mathbf{E}} = \mathbf{E}' / \mathbf{E}$
    \item \texttt{e(Y\_hat)}: $N \times 1$ matrix  containing $\hat{\mathbf{Y}} = \mathbf{Y}' / \mathbf{Y}$
    \item \texttt{e(W\_hat)}: $N \times 1$ matrix  containing $\hat{\mathbf{W}} = \mathbf{W}' / \mathbf{W}$
    \item \texttt{e(Q\_hat)}: $N \times 1$ matrix  containing $\hat{\mathbf{Q}} = \mathbf{Q}' / \mathbf{Q}$
    \item \texttt{e(E\_prime)}: $N \times 1$ matrix containing counterfactual expenditure $\mathbf{E}'$
    \item \texttt{e(Y\_prime)}: $N \times 1$ matrix containing counterfactual income $\mathbf{Y}'$
\end{itemize}

\section{Examples}

All three examples use the data set \texttt{GE\_gravity2\_example\_data.dta}, which is available online.\footnote{\texttt{https://github.com/rolf-campos/ge\_gravity2/raw/main/examples/ge\_gravity2\_example\_data.dta}} In this data set the string variable \texttt{iso\_o}, identifies the country of origin and the string variable \texttt{iso\_d}, identifies the country of destination. The variable \texttt{flow} contains bilateral trade flows from \texttt{iso\_o} to \texttt{iso\_d} and the variable \texttt{year} identifies years (in ten year increments).

\subsection{A simulation of the ex-ante effect of a trade agreement}
In a first example we simulate the effect of a free trade agreement. Suppose that we want to compute the expected effects of the North American Free Trade Agreement (NAFTA), a trade agreement signed by Canada, Mexico, and the United States, that entered into force in 1994. Also suppose that we expect the partial equilibrium effect of this trade agreement to be 0.5, meaning that countries that are part of this trade agreement are expected to increase their trade flows with other members by $\exp(0.5)-1 \approx 65\%$, if nothing else changes. The general equilibrium impact of this agreement can then be computed issuing the following commands in Stata.

\begin{stlog}
. use ge_gravity2_example_data.dta
. local beta = 0.500
{\smallskip}
. 
. ** Generate an indicator of NAFTA
. gen nafta = 0
{\smallskip}
. replace nafta = 1 if iso_o == "CAN" \& (iso_d == "MEX" | iso_d == "USA")
(12 real changes made)
{\smallskip}
. replace nafta = 1 if iso_o == "MEX" \& (iso_d == "CAN" | iso_d == "USA")
(12 real changes made)
{\smallskip}
. replace nafta = 1 if iso_o == "USA" \& (iso_d == "CAN" | iso_d == "MEX")
(12 real changes made)
{\smallskip}
. 
. ** Generate the partial equilibrium effect of NAFTA
. gen partial_effect = `beta' * nafta
{\smallskip}
. 
. ** Obtain general equilibrium effects of NAFTA using data for the year 1990.
. ** Report a table with results, and generate variables with counterfactual flows and W_hat.
. ** We use a trade elasticity of 5.03 and a supply elasticity of 1.24.
. 
.ge_gravity2 iso_o iso_d flow partial_effect if year==1990, theta(5.03) psi(1.24) results 
gen_X(flow_nafta) gen_w(welfare)
sorting...
solving...
solved!
{\smallskip}\nullskip
\end{stlog} 

After running this code, the command will print out a table that exhibits the change in trade flows, output, and welfare for all countries in the sample. Below, we show an excerpt from the table showing the first three lines and the countries involved in the trade agreement.

\begin{stlog}
{\smallskip}
                    Results for the prototypical trade model (percent changes)
{\smallskip}
{\TLC}\HLI{14}{\TOPT}\HLI{66}{\TRC}
{\VBAR}              {\VBAR}   Exports    Imports  IntlTrade   Domestic     Output    Welfare {\VBAR}
{\LFTT}\HLI{14}{\PLUS}\HLI{66}{\RGTT}
{\VBAR}          ARG {\VBAR}     0.066     -0.312     -0.020     -0.012     -0.002     -0.021 {\VBAR}
{\VBAR}          AUS {\VBAR}    -0.121     -0.127     -0.124      0.010     -0.005     -0.009 {\VBAR}
{\VBAR}          AUT {\VBAR}    -0.042     -0.008     -0.024      0.011     -0.001      0.006 {\VBAR}
\begin{center}
    (output omitted) 
\end{center}
{\VBAR}          CAN {\VBAR}    38.065     42.002     40.006     -5.476      2.544      4.662 {\VBAR}
\begin{center}
    (output omitted) 
\end{center}
{\VBAR}          MEX {\VBAR}    37.509     48.714     42.518     -3.632      1.698      3.182 {\VBAR}
\begin{center}
    (output omitted) 
\end{center}
{\VBAR}          USA {\VBAR}    16.074     13.065     14.397     -0.644      0.286      0.516 {\VBAR}
\begin{center}
    (output omitted) 
\end{center}
{\BLC}\HLI{14}{\BOTT}\HLI{66}{\BRC}
{\smallskip}\nullskip
\end{stlog} 

Because we have used the option \texttt{gen\_w(Welfare)}, the command will also have generated a new variable called \texttt{Welfare} that contains the comparative statics for welfare $\hat{W}_i = W'_i//W_i$ for every country $i$ ($i$ in this case refers to the country that is listed in the variable \texttt{iso\_o}). This new variable will have non-missing values only in the year 1990, because the command was issued specifying the condition \texttt{if year==1990}. Because the trade model is static, a researcher will want to use data only from a particular year, as we did in this example, except in rare cases.

\subsection{A quantification of the impact of time-varying border effects on welfare}

We now turn to a more complicated example that uses several years of data. In this example, we use estimates by \citet{CamposReggioTimini:23} of Spain's ``border thickness''. In their article, they measure the border thickness (a measure of how difficult it is to trade internationally) of Spain during the Franco regime and compare it to the border thickness of a synthetic control for Spain that is calculated as the average of other countries. They report estimates of the welfare loss implied by Spain's differential border thickness and interpret this welfare loss as the effect economic policies followed by Spain in the post-war period until 1975.

In this example we uses the same database as before and calculate the welfare gain that Spain would have experienced if it had had the border thickness of its synthetically constructed counterfactual. We perform the calculation for several years. We use a trade elasticity of 4 and three different values for the supply elasticity: 0, 1, and 2. A trade elasticity of 4 and a supply elasticity of zero corresponds to the original simulations by \citet{CamposReggioTimini:23}.

To avoid having to use a loop to simulate the model for each year, we make use of the fact that the \texttt{ge\_gravity2} command can be used with the \texttt{by} prefix. Although the identity and number of exporters/importers can vary across years, it is important that the data are square in each year (i.e., that there are the same exporters as importers, and no missing values).

\begin{stlog}
. use GE_gravity2_example_data.dta
{\smallskip}
. keep if year <= 1980
(12,500 observations deleted)
{\smallskip}
. ** Use the estimates of the partial equilibrium effect of Spain's border thickness
. // Source: Campos, R. G., Reggio, I., and Timini, J.,
. //         "Autarky in Franco's Spain: The costs of a closed economy",
. //         Economic History Review, 76 (2023), pp. 1259-1280.
. //         https://doi.org/10.1111/ehr.13243
. //         Taken from the replication materials for Figure 5.

. gen beta_Spain = .
(23,263 missing values generated)
{\smallskip}
. replace beta_Spain = -1.254 if year == 1950
(5,476 real changes made)
{\smallskip}
. replace beta_Spain = -0.937 if year == 1960
(5,929 real changes made)
{\smallskip}
. replace beta_Spain = -0.604 if year == 1970
(5,929 real changes made)
{\smallskip}
. replace beta_Spain = -0.694 if year == 1980 
(5,929 real changes made)
{\smallskip} 

. gen beta_Synthetic_Spain = .
(23,263 missing values generated)
{\smallskip}
. replace beta_Synthetic_Spain = -0.665 if year == 1950
(5,476 real changes made)
{\smallskip}
. replace beta_Synthetic_Spain = -0.438 if year == 1960
(5,929 real changes made)
{\smallskip}
. replace beta_Synthetic_Spain = -0.428 if year == 1970
(5,929 real changes made)
{\smallskip}
. replace beta_Synthetic_Spain = -0.653 if year == 1980 
(5,929 real changes made)
{\smallskip}
. 
. ** Generate an indicator of Spain's borders
. gen border = (iso_o != iso_d)
{\smallskip}
. gen border_Spain = border * (iso_o == "ESP" | iso_d == "ESP")
{\smallskip}

. ** Generate the partial equilibrium effect of the difference of Spain's actual border thickness relative to that of Synthetic Spain
. gen partial_effect = (beta_Synthetic_Spain - beta_Spain) * (border_Spain)
{\smallskip}

. ** Obtain the general equilibrium effect for different values of the supply elasticity
. ** Campos, Reggio, and Timini (2023) use a trade elasticity of 4.
. ** They (implicitly) use a supply elasticity of zero.
. ** We use the by prefix to calculate welfare for all years.
. bys year: ge_gravity2 iso_o iso_d flow partial_effect, theta(4) psi(0) gen_w(W0)
{\smallskip}
. bys year: ge_gravity2 iso_o iso_d flow partial_effect, theta(4) psi(1) gen_w(W1)
{\smallskip}
. bys year: ge_gravity2 iso_o iso_d flow partial_effect, theta(4) psi(2) gen_w(W2)
{\smallskip}

. ** Collapse and re-express in percentage points
. collapse (first) W0 W1 W2, by(iso_o year)
{\smallskip}
. replace W0 = 100 * (W0 - 1)
(305 real changes made)
{\smallskip}
. replace W1 = 100 * (W1 - 1)
(305 real changes made)
{\smallskip}
. replace W2 = 100 * (W2 - 1)
(305 real changes made)
{\smallskip}

. ** List welfare losses for Spain
. list if iso_o == "ESP"
{\smallskip}
     {\TLC}\HLI{50}{\TRC}
     {\VBAR} iso_o   year          W0          W1          W2 {\VBAR}
     {\LFTT}\HLI{50}{\RGTT}
 92. {\VBAR}   ESP   1950   -.7740736   -1.535153   -2.283889 {\VBAR}
 93. {\VBAR}   ESP   1960   -.6785989   -1.330477   -1.974016 {\VBAR}
 94. {\VBAR}   ESP   1970   -.3882885   -.6982863   -.9990215 {\VBAR}
 95. {\VBAR}   ESP   1980   -.1280367   -.2352655   -.3385067 {\VBAR}
     {\BLC}\HLI{50}{\BRC}
{\smallskip}\nullskip
\end{stlog} 
The results in this final listing show how the welfare calculations depend on different values of the supply elasticity.

\subsection{A simulation of a change in productivity}
The command \texttt{ge\_gravity2} can also be used to run simulations in which a country's productivity changes while its trade costs do not. In this last example, we simulate the impact that raising China's productivity by 10\% has on trade flows, output, and welfare of all countries in the world. We do this using the \texttt{a\_hat} option of the command.

\begin{stlog}
. use GE_gravity2_example_data.dta
{\smallskip}
 
. ** Perform a dry run to see how in what position China is ordered
. gen partial_effect = 0
{\smallskip}
. ge_gravity2 iso_o iso_d flow partial_effect if year == 1990, theta(5.03) psi(1.24)
sorting...
solving...
solved!
{\smallskip}
. matrix list e(W_hat)  // China is in position 11
{\smallskip}

. ** Generate the matrix with the productivity increase
. matrix A_hat = J(`e(N)', 1, 1)  // Matrix of size N x 1 filled with ones
{\smallskip}
. matrix A_hat[11, 1] = 1.1  // China's productivity increases by 10\%
{\smallskip}

. ** Run the command with the a_hat option
. ge_gravity2 iso_o iso_d flow partial_effect if year == 1990, theta(5.03) psi(1.24) a_hat(A_hat) results
sorting...
solving...
Using custom a_hat.
solved!\nullskip
\end{stlog} 

Below, we show an excerpt from the resulting table showing the first three lines and the line for China.

\begin{stlog}
{\smallskip}
                    Results for the prototypical trade model (percent changes)
{\smallskip}
{\TLC}\HLI{14}{\TOPT}\HLI{66}{\TRC}
{\VBAR}              {\VBAR}   Exports    Imports  IntlTrade   Domestic     Output    Welfare {\VBAR}
{\LFTT}\HLI{14}{\PLUS}\HLI{66}{\RGTT}
{\VBAR}          ARG {\VBAR}     0.109      0.029      0.091     -0.011      0.000     -0.010 {\VBAR}
{\VBAR}          AUS {\VBAR}     0.130      0.134      0.132     -0.011      0.005      0.009 {\VBAR}
{\VBAR}          AUT {\VBAR}     0.021      0.041      0.031     -0.001      0.003      0.011 {\VBAR}
\begin{center}
    (output omitted) 
\end{center}
{\VBAR}          CHN {\VBAR}     4.609      7.269      5.575     11.439      9.838     10.776 {\VBAR}
\begin{center}
    (output omitted) 
\end{center}
{\BLC}\HLI{14}{\BOTT}\HLI{66}{\BRC}
{\smallskip}

\nullskip
\end{stlog}

\section{Conclusion}
This paper introduces the command \texttt{ge\_gravity2}, which can be used in academic research and for policy analysis. By extending the capabilities of the original \texttt{ge\_gravity} command to models with positive aggregate supply elasticity, researchers and policymakers now have a powerful tool for analyzing the effects of trade policies on trade flows and welfare in the class of universal gravity models. The ease of use and seamless transition from \texttt{ge\_gravity} to \texttt{ge\_gravity2} make it straightforward to add this command to the toolkit of those working in general equilibrium trade modeling. 

We show how to derive the system of equations to compute comparative statics in a universal gravity model. We describe the algorithm that solves this nonlinear system of equations. We also present a prototypical trade model with positive supply elasticity, and show how it falls in the class of universal gravity models, and how the outputs of the command can be interpreted in the context of this model. 

We provide a detailed overview of the command's features, options, and potential applications, by going over three examples. The command's features allow to solve a wide range of economic geography models within the universal gravity framework.

\section{Acknowledgments}
We have greatly benefited from comments by Tom Zylkin. The views expressed in this paper are those of the authors and do therefore not necessarily reflect those of the Banco de Espa\~na or the Eurosystem.

\section{Programs and supplemental material}

To install the software files as they existed at the time of publication of this article type

\begin{stlog}
. ssc install ge_gravity2
\end{stlog} 
 



\bibliographystyle{sj}
\bibliography{ge_gravity2}

\Appendix
\section{Appendix: detailed derivations of results}
\subsection{Output in the prototypical trade model}
Gross production at each location $i$ is a Cobb-Douglas function. We specify this function in a more general form, by including an additional productivity term $Z_i$:
\begin{equation*}
    Q_i = Z_i (A_i L_i)^\zeta I_i^{1-\zeta},
\end{equation*}
where $\zeta \in (0,1]$ is the labor share, $Z_i$ is something akin to total factor productivity (taking intermediates to be a factor of production that is reproducible), $A_i$ is labor productivity, and $I_i$ is an intermediate input which is a constant elasticity of substitution (CES) aggregate of all varieties with the same elasticity of substitution as the bundle of consumption:
\begin{equation*} 
    I_i = \left(\sum_i \alpha_{ij}^\frac{1}{\sigma} x_{ij}^\frac{\sigma-1}{\sigma}\right)^\frac{\sigma}{\sigma-1},
\end{equation*}
where the parameters $\alpha_{ij} \geq 0$ are proportional to the trade costs $\tau_{ij}$.

The problem is separable and the cost function for intermediate goods can be derived as a first step before considering the cost-minimizing choice between labor and the bundle of intermediate goods. The cost-minimization problem for intermediate goods at the destination $j$ is:
\begin{equation*}
    \min_{x_{ij} \geq 0} \sum_{i} p_{ij} x_{ij} ,
\end{equation*}
subject to
\begin{equation*} 
    \left(\sum_i \alpha_{ij}^\frac{1}{\sigma} x_{ij}^\frac{\sigma-1}{\sigma}\right)^\frac{\sigma}{\sigma-1} \geq I_j
\end{equation*}

For the well-known CES production function, cost minimization implies a cost function of the form
\begin{equation*}
    C(I_j) = \left(\sum_{i} \alpha_{ij} p_{ij}^{1-\sigma}\right)^\frac{1}{1-\sigma} I_j = P_j I_j,
\end{equation*}
with price index
\begin{equation*}
    P_j = \sum_{i} (\alpha_{ij} p_{ij}^{1-\sigma})^\frac{1}{1-\sigma}.
\end{equation*}

The second stage of cost minimization (taking into account that intermediates have been chosen optimally) is:
\begin{equation*}
    \min_{\ell_j, I_j} = w_j \ell_j + P_j I_j,
\end{equation*}
subject to
\begin{equation*}
    Z_j (A_j \ell_j)^\zeta I_j^{1-\zeta} \geq Q_j 
\end{equation*}
The first order condition for $\ell$ is:
\begin{equation*}
    w_j = \lambda \zeta p_j Z_j A_j^\zeta \left(\frac{I_j}{\ell_j}\right)^{1-\zeta}.
\end{equation*}
The first order condition for $I_j$ is:
\begin{equation*}
   P_j = \lambda (1-\zeta) p_j Z_j A_j^\zeta \left(\frac{I_j}{\ell_j}\right)^{-\zeta}
\end{equation*}
Taking the ratio of these two equations, the optimal ratio of intermediates to labor is
\begin{equation*} 
    \frac{I_j}{\ell_j} = \frac{1-\zeta}{\zeta} \frac{w_j}{P_j}
\end{equation*}
This equation equates the marginal rate of transformation (MRT) to the input price ratio.

We now use this result in the production function to solve for $\ell$ as a function of $Q$:
\begin{align*}
   Q_j &= Z_j (A_j \ell_j)^\zeta I_i^{1-\zeta} \\
   &= \ell_j Z_j (A_j)^\zeta \left(\frac{I_i}{\ell_j}\right)^{1-\zeta} \\
   &= \ell_j A_j (Z_j)^\zeta \left(\frac{1-\zeta}{\zeta} \frac{w_j}{P_j}\right)^{1-\zeta}, \\
   \ell_j &= (A_j)^{-\zeta} \left(\frac{1-\zeta}{\zeta} \frac{w_j}{P_j}\right)^{\zeta-1} \frac{Q_j}{Z_j}. 
\end{align*}
Using the equation that equates the MRT to the input price ratio one more time, we can also solve for $I$:
\begin{equation*}
    I_j = (A_j)^{-\zeta} \left(\frac{1-\zeta}{\zeta} \frac{w_j}{P_j}\right)^{\zeta} \frac{Q_j}{Z_j}.
\end{equation*}

The cost function can be obtained by substituting the conditional factor demands that we obtained into the expenditure on labor input and the intermediate input:
\begin{align*}
    C_j &= w_j \ell_j + P_j I_j \\
    &= w_j (A_j)^{-\zeta} \left(\frac{1-\zeta}{\zeta} \frac{w_j}{P_j}\right)^{\zeta-1} \frac{Q_j}{Z_j} + P_j (A_j)^{-\zeta} \left(\frac{1-\zeta}{\zeta} \frac{w_j}{P_j}\right)^{\zeta} \frac{Q_j}{Z_j} \\
    &= \frac{Q_j}{Z_j} (A_j)^{-\zeta} \left[w_j  \left(\frac{1-\zeta}{\zeta} \frac{w_j}{P_j}\right)^{\zeta-1}  + P_j \left(\frac{1-\zeta}{\zeta} \frac{w_j}{P_j}\right)^{\zeta} \right] \nonumber \\
    &= \frac{Q_j}{Z_j} (A_j)^{-\zeta} \left[\left(\frac{1-\zeta}{\zeta} \right)^{\zeta-1} w_j^\zeta P_j^{1-\zeta} + \left(\frac{1-\zeta}{\zeta} \right)^{\zeta}  w_j^\zeta P_j^{1-\zeta} \right] \nonumber \\
    &= \frac{Q_j}{Z_j} (A_j)^{-\zeta} w_j^\zeta P_j^{1-\zeta} \left[\left(\frac{1-\zeta}{\zeta} \right)^{\zeta-1} + \left(\frac{1-\zeta}{\zeta} \right)^{\zeta}  \right] \nonumber \\
    &= \frac{Q_j}{Z_j} B(\zeta) A_j^{-\zeta} w_j^\zeta P_j^{1-\zeta}
\end{align*}

Because of constant returns to scale, the cost function is linear in output, and quantities would be zero or infinity unless the output price $p_i$ at which the good produced is sold at each location $i$ is exactly
\begin{equation*}
    p_i = (B(\zeta)/Z_i) (w_i/A_i)^\zeta P_i^{1-\zeta},
\end{equation*}
where $w_i$ is the wage and $P_i$ is the CES price index (this also implies that profits will be zero, as is usual in the case with constant returns to scale). Solving for $w_i/p_i$, in equilibrium, the ratio of wages to output prices is a function of the ``real price'' $p_i/P_i$, with elasticity $\psi \equiv \frac{1-\zeta}{\zeta}$:
\begin{equation*}
    \frac{w_i}{p_i} = A_i \left(\frac{Z_i}{B(\zeta)}\right)^\frac{1}{\zeta} \left(\frac{p_i}{P_i}\right)^\frac{1-\zeta}{\zeta}
\end{equation*}
Evaluating the equation of the MRT at $\ell_i = L_i$ and substituting into this relation the result for $w_i/p_i$:
\begin{align*}
    I_i &= \frac{1-\zeta}{\zeta} \frac{w_i}{P_i} L_i \nonumber \\
    &= \frac{1-\zeta}{\zeta} \frac{w_i}{p_i} \frac{p_i}{P_i} L_i \nonumber \\
    &= \frac{1-\zeta}{\zeta}  A_i \left(\frac{Z_i}{B(\zeta)}\right)^\frac{1}{\zeta} \left(\frac{p_i}{P_i}\right)^\frac{1-\zeta}{\zeta} \frac{p_i}{P_i} L_i \nonumber \\
    &= \frac{1-\zeta}{\zeta} A_i L_i \left(\frac{Z_i}{B(\zeta)}\right)^\frac{1}{\zeta} \left(\frac{p_i}{P_i}\right)^\frac{1}{\zeta}
\end{align*}
Substituting this expression and $\ell_i = L_i$ into the production function:
\begin{align*}
    Q_i &= Z_i (A_i L_i)^\zeta I_i^{1-\zeta} \nonumber \\
    &= Z_i(A_i L_i)^\zeta \left(  \frac{1-\zeta}{\zeta} A_i L_i \left(\frac{Z_i}{B(\zeta)}\right)^\frac{1}{\zeta} \left(\frac{p_i}{P_i}\right)^\frac{1}{\zeta} \right)^{1-\zeta} \nonumber \\
    &= Z_i \left(  \frac{1-\zeta}{\zeta} \left(\frac{Z_i}{B(\zeta)}\right)^\frac{1}{\zeta} \right)^{1-\zeta} A_i L_i \left(\frac{p_i}{P_i}\right)^\frac{1-\zeta}{\zeta} \nonumber \\
    &= K(\zeta) A_i L_i Z_i^{1+\psi}\left(\frac{p_i}{P_i}\right)^\psi,
\end{align*}
where $\psi$ is the supply elasticity.


\subsection{Welfare in the prototypical trade model}

The utility function is 
\begin{equation*}
    U_j = \left(\sum_i \alpha_{ij}^\frac{1}{\sigma} q_{ij}^\frac{\sigma-1}{\sigma}\right)^\frac{\sigma}{\sigma-1}.
\end{equation*}
The parameter $\sigma > 1$ in the utility function is the elasticity of substitution between varieties $q_{ij} \geq 0$. The parameters $\alpha_{ij} \geq 0$ are demand shifters that are proportional to the bilateral trade costs $\tau_{ij}$. 
The budget constraint is given by
\begin{equation*}
    \sum_i p_{ij} q_{ij} = E_j,
\end{equation*}
where $p_{ij}$ is the price paid at location $j$ for the good $q_{ij}$ and $E_j$ total expenditure at location $j$.

Utility maximization with CES preferences is well-known and details are available in many textbooks. We directly state the results for this class of preferences. Utility maximization in this case implies an optimal demand function of the form
\begin{equation*}
    q_{ij}^\ast = \alpha_{ij}\left(\frac{p_{ij}}{P_j}\right)^{-\sigma} \frac{E_j}{P_j},
\end{equation*}
with the price index $P_j$ defined by
\begin{equation*}
    P_j = \sum_{i} (\alpha_{ij} p_{ij}^{1-\sigma})^\frac{1}{1-\sigma}.
\end{equation*}
This price index is the same as the price index for intermediate goods used in production.

By substituting optimal demand functions in the utility function, we obtain the per-capita indirect utility function:
\begin{align*}
    W_j \equiv \frac{U_j(\{q_{ij}^\ast\})}{L_j} &= \left(\sum_i \alpha_{ij}^\frac{1}{\sigma} \left[\alpha_{ij}\left(\frac{p_{ij}}{P_j}\right)^{-\sigma} \frac{E_j}{P_j}\right]^\frac{\sigma-1}{\sigma}\right)^\frac{\sigma}{\sigma-1} \frac{1}{L_j} \nonumber \\
    &= \left(\sum_i \alpha_{ij}\left[\left(\frac{p_{ij}}{P_j}\right)^{-\sigma} \right]^\frac{\sigma-1}{\sigma}\right)^\frac{\sigma}{\sigma-1} \frac{E_j}{P_j} \frac{1}{L_j} \\
    &=  \left(P_j^{\sigma-1} \sum_i \alpha_{ij} p_{ij}^{1-\sigma}\right)^\frac{\sigma}{\sigma-1}  \frac{\Xi \xi_j w_j L_j}{P_j} \frac{1}{L_j}  \\
    &= \Xi \xi_j \frac{w_j}{P_j}
\end{align*}
Welfare in this model is given by the real wage (in terms of the price index $P_j$, not the producer price $p_j$). Using the results from the production side of the model, this real wage can be expressed as
\begin{equation*}
    \frac{w_j}{P_j} = \left(\frac{w_j}{p_j}\right) \left(\frac{p_j}{P_j}\right)  = (Z_j/B(\zeta))^\frac{1}{\zeta} A_j \left(\frac{p_j}{P_j}\right)^{\frac{1-\zeta}{\zeta}} \left(\frac{p_j}{P_j}\right) = (Z_j/B(\zeta))^\frac{1}{\zeta} A_j \left(\frac{p_j}{P_j}\right)^{\frac{1}{\zeta}}.
\end{equation*}
If $\psi = (1-\zeta)/\zeta$, then $1/\zeta = 1+\psi$, so this last expression can be written as:
\begin{equation*}
    \frac{w_j}{P_j} = (Z_j/B(\psi))^{1+\psi} A_j \left(\frac{p_j}{P_j}\right)^{1+\psi}.
\end{equation*}

\subsection{Summary of results for the prototypical trade model}
The real wage in terms or the output price ($w/p$), the real wage in terms of the consumption price ($w/P$), output ($Q$), income ($Y$) and welfare ($W$) satisfy the following proportionality relations:

\begin{align*}
    \frac{w_i}{p_i} &\propto A_i Z_i^{1+\psi} \left(\frac{p_i}{P_i}\right)^{\psi}, \nonumber \\
    Q_i &\propto A_i L_i Z_i^{1+\psi} \left(\frac{p_i}{P_i}\right)^{\psi}, \nonumber \\
    Y_i = p_i Q_i  &\propto A_i L_i Z_i^{1+\psi} \frac{p_i^{1+\psi}}{P_i^{\psi}}, \nonumber \\
    W_i = \Xi \xi_i \frac{w_i}{P_i} &\propto \Xi \xi_i A_i \left(Z_i\frac{p_i}{P_i}\right)^{1+\psi}.
\end{align*}

We notice that output per capita ($Q_i/L_i$) and the real wage in terms of the output price are also proportional to each other. In terms of the notation used for the universal gravity framework, we have $\overline{c}_i = A_i L_i$. Finally, if $L_i$ and $Z_i$ are constant, as in the prototypical trade model, then the proportionality relations stated in the main text are obtained.

\subsection{Derivation of the system of equations for equilibrium prices with unbalanced trade}

We start out by inserting the international arbitrage condition stated as Property~1 in the equation for trade flows implied by Property~2 (CES demand). This yields
\begin{equation*} 
    X_{ij} = (\tau_{ij} p_i)^{-\theta} P_j^\theta E_j.
\end{equation*}
Next, we use Property~4 (output market clearing), and multiply both sides by $p_i$ to obtain:
\begin{align*} 
    Y_i \equiv p_i Q_i &= p_i \sum_j \tau_{ij} q_{ij} \nonumber \\
    &= \sum_j p_{ij} q_{ij} \nonumber \\
    &= \sum_j X_{ij} \nonumber \\
    &= \sum_j (\tau_{ij} p_i)^{-\theta} P_j^\theta E_j \nonumber \\
    &= \sum_j (\tau_{ij} p_i)^{-\theta} P_j^\theta \Xi \xi_j p_j Q_j
\end{align*}
The steps involved in the equations above are as follows. The second line uses Property~1. The third line uses the definition of $X_{ij}$. The fourth line uses the equation we have derived before. The fifth line uses Property~5.

Now, we can substitute output using Property~3:
\begin{equation*}
    p_i \left(\kappa\overline{c}_i \left( \frac{p_i}{P_i}\right)^\psi\right)= \sum_j (\tau_{ij} p_i)^{-\theta} P_j^\theta \Xi \xi_j p_j \left(\kappa\overline{c}_j \left( \frac{p_j}{P_j}\right)^\psi\right),
\end{equation*}
or
\begin{equation*} 
    p_i^{1+\theta} \left(\overline{c}_i \left( \frac{p_i}{P_i}\right)^\psi\right)= \sum_j \tau_{ij}^{-\theta} P_j^\theta \Xi \xi_j p_j \left(\overline{c}_j \left( \frac{p_j}{P_j}\right)^\psi\right).
\end{equation*}
This equation holds for all $i$. 

Combining the price index with Property~1 delivers:
\begin{equation*}
    P_j^{-\theta} = \sum_{i} \tau_{ij}^{-\theta} p_{i}^{-\theta}.
\end{equation*} 
This equation holds for all $i$. 

There are a total of $2N$ equations for $2N$ prices. Because these equations are homogeneous of degree zero, they determine prices only up to scale. Property~6 normalizes the scale.

\subsection{Comparative statics for prices}
To derive comparative statics for prices, we start by taking the ratio of the first equation in the system of equations that determines prices evaluated in the counterfactual and in the baseline:
\begin{equation*} 
    \hat{p}_i^{1+\theta+\psi} \hat{P}_i^{-\psi} \left(\frac{\overline{c}'_i}{\overline{c}_i}\right) = \frac{\sum_j (\tau'_{ij})^{-\theta} (P'_j)^\theta \Xi' \xi'_j p'_j \left(\overline{c}'_j \left( \frac{p'_j}{P'_j}\right)^\psi\right)}{\sum_j \tau_{ij}^{-\theta} P_j^\theta \Xi \xi_j p_j \left(\overline{c}_j \left( \frac{p_j}{P_j}\right)^\psi\right)}
\end{equation*}

The denominator of the expression on the right hand side can be simplified as follows: 
\begin{align*}
    \sum_j \tau_{ij}^{-\theta} P_j^\theta \Xi \xi_j p_j \left(\overline{c}_j \left( \frac{p_j}{P_j}\right)^\psi\right) &= p_i^\theta \sum_j (\tau_{ij} p_i)^{-\theta} P_j^\theta \Xi \xi_j p_j \left(\overline{c}_j \left( \frac{p_j}{P_j}\right)^\psi\right) \nonumber \\
    &= p_i^\theta \sum_j (\tau_{ij} p_i)^{-\theta} P_j^\theta \Xi \xi_j  p_j Q_j \nonumber \\
    & = p_i^\theta Y_i.
\end{align*}
The first equality above follows from multiplying and dividing by $p_i^\theta$, the second is obtained by replacing $\overline{c}_j \left( \frac{p_j}{P_j}^\psi \right)$ with $p_j Q_j$, and the last equality follows from the equation $Y_i=\sum_j (\tau_{ij} p_i)^{-\theta} P_j^\theta \Xi \xi_j p_j Q_j$ that was shown in the derivation of the system of equations for prices.


Now we substitute this expression for the denominator into the original equation:
\begin{equation*} 
    \hat{p}_i^{1+\theta+\psi} \hat{P}_i^{-\psi}\left(\frac{\overline{c}'_i}{\overline{c}_i}\right) = \frac{\sum_j (\tau'_{ij})^{-\theta} (P'_j)^\theta \Xi' \xi'_j p'_j \left(\overline{c}'_j \left( \frac{p'_j}{P'_j}\right)^\psi\right)}{p_i^\theta Y_i}.
\end{equation*}
The next step is to convert the variables with the apostrophe in the numerator of the expression on the right hand side to ``hat'' variables, i.e., using the replacement $x' = \hat{x} x$ whenever possible. This leads to
\begin{align*}
    \frac{\sum_j (\tau'_{ij})^{-\theta} (P'_j)^\theta \Xi' \xi'_j p'_j \left(\overline{c}'_j \left( \frac{p'_j}{P'_j}\right)^\psi\right)}{p_i^\theta Y_i} &=  \frac{\sum_j (\hat{\tau}_{ij}\tau_{ij})^{-\theta} (\hat{P}_j P_j)^\theta \hat{\Xi}\Xi \hat{\xi}_j \xi_j \hat{p}_j p_j \hat{c}_j\overline{c}_j \left( \frac{\hat{p}_j}{\hat{P}_j} \right)^\psi \left( \frac{p_j}{P_j} \right)^\psi}{p_i^\theta Y_i} \nonumber \\
    &= \frac{\sum_j (\hat{\tau}_{ij})^{-\theta} (\hat{P}_j)^\theta \hat{\Xi}\hat{\xi}_j \hat{p}_j \hat{c}_j \left( \frac{\hat{p}_j}{\hat{P}_j} \right)^\psi \left[\tau_{ij}^{-\theta} P_j^\theta \Xi \xi_j p_j \left(\overline{c}_j \left( \frac{p_j}{P_j}\right)^\psi\right)\right]}{p_i^\theta Y_i} \nonumber \\
    &= \frac{\sum_j (\hat{\tau}_{ij})^{-\theta} (\hat{P}_j)^\theta \hat{\Xi}\hat{\xi}_j \hat{p}_j \hat{c}_j \left( \frac{\hat{p}_j}{\hat{P}_j} \right)^\psi \left[p_i^\theta X_{ij}\right]}{p_i^\theta Y_i} \nonumber \\
    &= \sum_j \left[\frac{X_{ij}}{Y_i}\right] (\hat{\tau}_{ij})^{-\theta} (\hat{P}_j)^\theta \hat{\Xi}\hat{\xi}_j \hat{p}_j \hat{c}_j \left( \frac{\hat{p}_j}{\hat{P}_j} \right)^\psi 
\end{align*}
Combining the left hand side with the right hand side we obtain the final version of the first equation that can be used to solve for comparative statics in prices:
\begin{equation*}
    \hat{p}_i^{1+\theta+\psi} \hat{P}_i^{-\psi} \hat{c}_i = \hat{\Xi} \sum_j \left[\frac{X_{ij}}{Y_i}\right] (\hat{\tau}_{ij})^{-\theta} (\hat{P}_j)^\theta \hat{\xi}_j \hat{p}_j \hat{c}_j \left( \frac{\hat{p}_j}{\hat{P}_j} \right)^\psi.
\end{equation*}

To obtain the next equation, we take the ratio of the definition of the price index in the counterfactual and in the baseline (after switching $i$ and $j$ in the notation):
\begin{equation*} 
    \hat{P}_i^{-\theta} = \frac{(P'_i)^{-\theta}}{P_i^{-\theta}} = \frac{\sum_{j} (\tau'_{ji})^{-\theta} (p'_{j})^{-\theta}}{P_i^{-\theta}}
\end{equation*}

On the right hand side of this expression we perform the ``hat'' variable substitutions, as before:
\begin{align*}
    \frac{\sum_{j} (\tau')_{ji}^{-\theta} (p')_{j}^{-\theta}}{P_i^{-\theta}} &= \frac{\sum_{j} \hat{\tau}_{ji}^{-\theta} \hat{p}_{j}^{-\theta} \tau_{ji}^{-\theta} p_{j}^{-\theta}}{P_i^{-\theta}} \nonumber \\
    &= \sum_{j} \hat{\tau}_{ji}^{-\theta} \hat{p}_{j}^{-\theta} \left[\frac{ \tau_{ji}^{-\theta} p_{j}^{-\theta}}{P_i^{-\theta}}\right] \nonumber \\
    &= \sum_{j} \hat{\tau}_{ji}^{-\theta} \hat{p}_{j}^{-\theta} \left[ \frac{P_i^{-\theta} \frac{1}{E_i} \tau_{ji}^{-\theta} p_{j}^{-\theta} P_i^{\theta} E_i}{P_i^{-\theta}} \right] \nonumber \\
    &= \sum_{j} \hat{\tau}_{ji}^{-\theta} \hat{p}_{j}^{-\theta} \left[ \frac{X_{ji}}{E_i} \right] \nonumber \\
\end{align*}
By combining the left hand side and the right hand side, we get the final version of the second equation that can be used to solve for comparative statics in prices:
\begin{equation*}
    \hat{P}_i^{-\theta} = \sum_{j} \left[ \frac{X_{ji}}{E_i} \right] \hat{\tau}_{ji}^{-\theta} \hat{p}_{j}^{-\theta}.
\end{equation*}

The only endogenous element needed for the computation of prices that still needs to be solved for is $\hat{\Xi}$. We start from the following equalities:
\begin{equation*}
    \overline{Y} = \sum_{i=1}^N Y'_i = \sum_{i=1}^N E'_i = \Xi' \sum_{i=1}^N \xi'_i Y'_i = \hat{\Xi} \sum_{i=1}^N \hat{\xi}_i \hat{Y}_i \Xi \xi_i Y_i = \hat{\Xi} \sum_{i=1}^N \hat{\xi}_i \hat{Y}_i E_i.
\end{equation*}
The first equality follows from Property~6 applied to the counterfactual equilibrium. The second equality follows from aggregate consistency in the counterfactual equilibrium, which is implied by Property~5. The third equality follows from the definition of $\Xi$ given in Proposition~5 when it is applied to the counterfactual equilibrium. The fourth equality uses the definition of ``hat variables'' and the last equality uses Proposition~5, this time applied to the baseline equilibrium. Taking the first and the last expression in the previous chain of equalities yields the following expression for $\hat{\Xi}$:
\begin{equation*}
    \hat{\Xi} = \frac{\overline{Y}}{\sum_{i=1}^N \hat{\xi}_i \hat{Y}_i E_i}.
\end{equation*}
This expression depends on the endogenous variable $\hat{Y}_i$. By Property 3 we have
\begin{equation*}
    \hat{Q}_i = \hat{c}_i\left(\frac{\hat{p_i}}{\hat{P}_i}\right)^\psi.
\end{equation*}
Multiplying both sides by $\hat{p}_i$ we obtain
\begin{equation*}
    \hat{Y}_i = \hat{p}_i\hat{Q}_i = \hat{c}_i\frac{\hat{p_i}^{1+\psi}}{\hat{P}_i^\psi}.
\end{equation*}
Therefore, $\hat{\Xi}$ can be expressed in terms of price changes and known quantities as follows: 
\begin{equation*}
    \hat{\Xi} = \frac{\overline{Y}}{\sum_{i=1}^N \hat{\xi}_i \hat{c}_i\frac{\hat{p_i}^{1+\psi}}{\hat{P}_i^\psi} E_i} =\frac{1}{\sum_{i=1}^N \hat{\xi}_i \hat{c}_i\frac{\hat{p_i}^{1+\psi}}{\hat{P}_i^\psi} (E_i/\overline{Y})}.
\end{equation*}
\end{document}